\documentclass[aps,pra,twocolumn,showpacs,groupedaddress]{revtex4-1}
\usepackage[english]{babel}
\usepackage[T1]{fontenc}
\usepackage{natbib}
\usepackage{amssymb}
\usepackage{bm}
\usepackage{amsmath}
\usepackage[mathcal]{eucal}
\usepackage{color}
\usepackage{graphicx}
\usepackage{hyperref}

\newcommand {\nc} {\newcommand}
\nc {\beq} {\begin{eqnarray}}
\nc {\eeqn} [1] {\label{#1} \end{eqnarray}}
\nc {\eoln} [1] {\label{#1} \\}
\nc {\eol} {\nonumber \\}
\nc {\rref} [1] {(\ref{#1})}
\nc {\Eq} [1] {Eq.~(\ref{#1})}
\nc {\Ref} [1] {Ref.~\cite{#1}}
\nc {\la} {\mbox{$\langle$}}
\nc {\ra} {\mbox{$\rangle$}}
\nc {\dem} {\mbox{$\frac{1}{2}$}}
\nc {\cP} {\mathcal{P}}
\nc {\cN} {\mathcal{N}}
\nc {\ve} [1] {\mbox{\boldmath $#1$}}
\nc {\arrow} [2] {\mbox{$\mathop{\rightarrow}\limits_{#1 \rightarrow #2}$}}
\nc {\red}[1] {\textcolor{red}{#1}}

\begin{document}

\title{Relativistic polarizabilities with the Lagrange-mesh method}
\author{Livio Filippin}
\email[]{Livio.Filippin@ulb.ac.be}
\affiliation{Chimie Quantique et Photophysique, C.P.\ 160/09, Universit\'e Libre de Bruxelles (ULB), B-1050 Brussels, Belgium}
\author{Michel Godefroid}
\email[]{mrgodef@ulb.ac.be}
\affiliation{Chimie Quantique et Photophysique, C.P.\ 160/09, Universit\'e Libre de Bruxelles (ULB), B-1050 Brussels, Belgium}
\author{Daniel Baye}
\email[]{dbaye@ulb.ac.be}
\affiliation{Physique Quantique, C.P.\ 165/82, and Physique Nucl\'eaire Th\'eorique et Physique Math\'ematique, C.P.\ 229, Universit\'e Libre de Bruxelles (ULB), B-1050 Brussels, Belgium}
\date{\today}

\begin{abstract}
Relativistic dipolar to hexadecapolar polarizabilities of the ground state and some excited states 
of hydrogenic atoms are calculated by using numerically exact energies and wave functions 
obtained from the Dirac equation with the Lagrange-mesh method. 
This approach is an approximate variational method 
taking the form of equations on a grid because of the use of a Gauss quadrature approximation. 
The partial polarizabilities conserving the absolute value of the quantum number $\kappa$ 
are also numerically exact with small numbers of mesh points. 
The ones where $|\kappa|$ changes are very accurate when using three different meshes 
for the initial and final wave functions and for the calculation of matrix elements. 
The polarizabilities of the $n=2$ excited states of hydrogenic atoms are also studied 
with a separate treatment of the final states that are degenerate at the nonrelativistic approximation. 
The method provides high accuracies for polarizabilities of a particle in a Yukawa potential 
and is applied to a hydrogen atom embedded in a Debye plasma. 
\end{abstract}

\pacs{31.15.ap, 03.65.Pm, 32.10.Dk, 02.70.Hm}

\maketitle

\section{Introduction}

Atomic polarizabilities are very useful in various domains of physics \cite{MSC10}. 
They play a role in effects where an atom or ion can be deformed by the effect of a field. 
Well known examples are dielectric constants or refractive indexes. 
In some cases, experiments have reached such a high accuracy that relativistic effects must be precisely taken into account \cite{TZZ12}. 
The polarizabilities become a testing ground for accurate numerical methods. 
Here we show that highly accurate values can be obtained not only for the ground state of hydrogenic ions 
but also for excited states and for other types of central potentials. 

In the nonrelativistic case, static polarizabilities for spherical quantum numbers $nl$ provide exact limits of dynamical polarizabilities when the frequency tends to zero. 
They are also the limit of relativistic polarizabilities when the fine structure constant $\alpha$ is set to zero. 
For the hydrogen atom, analytical expressions have been derived first for $s$ states \cite{Bu37,Mc76} and later for all states \cite{KMM97,KMM05}. 
They can also be derived from exact numerical calculations \cite{Ba12}. 

In the relativistic case, exact static dipole polarizabilities are known only for the ground state \cite{Ya03,SM04} and the $2s$ excited state \cite{Ya03} of hydrogenic atoms. 
They are much more complicated than in the nonrelativistic case, 
involving $_3F_2$ hypergeometric functions. Series expansions in powers of $\alpha Z$ have also been established 
\cite{BP69,Ka77,ZMR72}. 
Accurate numerical values for the dipolar to hexadecapolar polarizabilities of the hydrogenic ground states are determined with B-splines for atomic numbers $Z = 1$ to 100 in \Ref{TZZ12}. 

In the present work, our aim is to calculate accurate numerical polarizabilities for various states described by the Dirac equation with the Lagrange-mesh method. 
The Lagrange-mesh method is an approximate variational method involving a basis of Lagrange functions related to a set of mesh points associated with a Gauss quadrature \cite{BH86,VMB93,Ba06}. 
Lagrange functions are continuous functions that vanish at all points of the corresponding mesh but one. 
The principal simplification appearing in the Lagrange-mesh method is that matrix elements are calculated with the Gauss quadrature. 
The potential matrix is then diagonal and only involves values of the potential at mesh points. 
This method has provided numerically exact (i.e.\ exact up to rounding errors) polarizabilities in the nonrelativistic hydrogenic case \cite{Ba12} because the basis is able to exactly reproduce the hydrogenic wave functions and the Gauss quadrature is exact for the relevant matrix elements with small numbers of mesh points. 
Recently, we have shown that numerically exact solutions of the Coulomb-Dirac equation can also be obtained with this method \cite{BFG14}. 
More generally, the method is very accurate for central potentials as illustrated with Yukawa potentials in \Ref{BFG14}. 
Here we use the obtained wave functions to study multipolar polarizabilities of the ground state and some excited states in the hydrogenic and Yukawa cases. 

In Sec.~\ref{sec:pola}, the nonrelativistic expressions of polarizabilities of a particle in a potential are recalled. 
The corresponding relativistic expressions are derived. 
In Sec.~\ref{sec:LMM}, the principle of the Lagrange-mesh method is summarized 
and relativistic polarizabilities are obtained with the associated Gauss quadrature. 
A technique involving three different meshes is established which gives better results. 
In Sec.~\ref{sec:res}, numerical results are presented for hydrogenic atoms 
and for a particle in Yukawa potentials. 
Section \ref{sec:conc} contains concluding remarks. 

For the fine-structure constant, we use the CODATA 2010 value $1/\alpha = 137.035 999 074$ \cite{MTN12}. 


\section{Nonrelativistic and relativistic polarizabilities}
\label{sec:pola}

\subsection{Nonrelativistic polarizabilities}
\label{sec:Schrodinger}

Before considering polarizabilities in a relativistic context, it is useful to summarize their nonrelativistic calculation. 

We consider the polarizability induced by the multipole operator 
\beq
C^{(\lambda)}_\mu (\Omega) r^\lambda = \sqrt{4\pi/(2\lambda+1)} Y^{(\lambda)}_\mu (\Omega) r^\lambda 
\eeqn{nr.0}
where $r$ is the radial coordinate and $\Omega$ represents the angular spherical coordinates. 
For a particle with mass $m$ in a potential $V(r)$, the radial functions $\psi_{nl}$ with radial or principal quantum number $n$ and orbital angular momentum quantum number $l$ are eigenfunctions of the radial Hamiltonian 
\beq
H_l = \frac{1}{2} \left[ -\frac{d^2}{dr^2} + \frac{l(l+1)}{r^2} \right] + V(r)
\eeqn{nr.1}
with eigenvalues $E_{nl}$, in units $\hbar = m = 1$. 

The polarizability of a level $nl$ for the $\lambda$-multipole operator is given by 
\beq
\alpha^{(nl)}_{\lambda} = \frac{1}{2\lambda+1} \sum_{l'=|l-\lambda|}^{l+\lambda}
\!\!\!\!^\prime\ \alpha^{(nll')}_{\lambda} 
\eeqn{nr.2}
where the prime means that the sum runs by steps of two. The reduced polarizabilities $\alpha^{(nll')}_{\lambda}$ appearing in this expression read 
\beq
\alpha^{(nll')}_{\lambda} & = & 2(2l'+1) 
\left ( \begin{array}{ccc} l' & \lambda & l \\  0 & 0 & 0 \end{array} \right )^2 \eol
& & \times \sum_{n'} \frac{[\int_0^\infty \psi_{n'l'}(r) r^\lambda \psi_{nl}(r) dr]^2}{E_{n'l'}-E_{nl}}.
\eeqn{nr.3}
In the Coulomb case, the term with principal quantum number $n' = n$ must be excluded. 
In this definition, the sum over $n'$ should be understood as representing a sum over discrete states and an integral over the continuum. 
A direct calculation is thus not easy. Hence it is useful to use a more compact expression. 

For partial wave $l$, the radial functions $\psi^{(1)}_{nll'}$ at the first order of perturbation theory are solutions of the inhomogeneous radial equations \cite{DL55,CT06,Ba12} 
\beq
\left( H_{l'} - E_{nl} \right) \psi^{(1)}_{nll'} (r) = (1 - \cP_{nl'}) r^\lambda \psi_{nl} (r)
\eeqn{nr.4}
where $\psi_{nl}$ is the radial wave function of the studied state. 
In the hydrogenic case, for $n>1$, the operator $\cP_{nl'}$ is a projector on the radial function of the $nl'$ state degenerate with the $nl$ state 
so that the function $\widetilde{\psi}^{(1)}_{nll'} = (1 - \cP_{nl'}) \psi^{(1)}_{nll'}$ does not contain the degenerate component. 
The reduced polarizabilities can be rewritten as \cite{DL55,Ba12} 
\beq
\alpha^{(nll')}_{\lambda} = 2(2l'+1) 
\left( \begin{array}{ccc} l' & \lambda & l \\  0 & 0 & 0 \end{array} \right)^2 
\int_0^\infty \widetilde{\psi}_{nll'}^{(1)}(r) r^\lambda \psi_{nl}(r) dr. \eol
\eeqn{nr.5}
They simply involve a single integral but require solving the inhomogeneous equation \rref{nr.4}. 
When present, the projection operator introduces a significant complication for numerical calculations. 

The equivalent equations \rref{nr.3} and \rref{nr.5} are not redundant; 
they are complementary. In the Coulomb case, analytical calculations are simpler with \Eq{nr.5}. 
Moreover, \Eq{nr.4} has an exact solution $\widetilde{\psi}_{nll'}^{(1)}$ 
which contains the exponential $\exp(-Zr/n)$ multiplied by a polynomial of degree $n + \lambda + 1$. 
This leads in \Eq{nr.5} to the exact values of the polarizabilities, a result not easy to obtain with the equivalent expression \rref{nr.3}. 

In numerical calculations, the respective merits of Eqs.~\rref{nr.3} and \rref{nr.5} become different. 
The sum-integral over $n'$ in \Eq{nr.3} is replaced by a finite sum over pseudostates where the optimal way of choosing these pseudostates is not obvious but the elimination of degenerate states in the Coulomb case is very easy. 
Equations \rref{nr.4} and \rref{nr.5} indicate that an exact polarizability can be obtained in the Coulomb case with pseudostates containing the same exponential $\exp(-Zr/n)$ multiplied by a polynomial. 
Equation \rref{nr.3} requires a diagonalization of the matrix corresponding to the final orbital momentum $l'$ in the pseudostate basis, 
while \Eq{nr.5} requires the solution of an algebraic system derived from \Eq{nr.4}, or the inversion of a matrix. 

With the Lagrange-mesh method described below, a striking property, not emphasized enough in \Ref{Ba12}, is that both approaches \rref{nr.3} and \rref{nr.5} lead for any level to the same exact results, up to rounding errors. 
Indeed, when the matrix representing the left-hand side of \Eq{nr.4} is inverted by using its spectral decomposition, 
the resulting expression is then identical to a pseudostate expansion based on \rref{nr.3} (see Appendix \ref{sec:A}). 
We now analyze the corresponding expressions in the relativistic case and show that the same ideas can be exploited but with some differences.  


\subsection{Relativistic polarizabilities}
\label{sec:Dirac}

In atomic units, the coupled radial Dirac equations read \cite{Gr07} 
\beq
H_\kappa \left( \begin{array}{c} P_{n\kappa}(r) \\ Q_{n\kappa}(r) \end{array} \right) 
= E_{n\kappa} \left( \begin{array}{c} P_{n\kappa}(r) \\ Q_{n\kappa}(r) \end{array} \right)
\eeqn{r.1}
with the Hamiltonian matrix 
\beq
H_\kappa = \left( \begin{array}{cc} 
V(r) & c \left( -\frac{d}{dr} + \frac{\kappa}{r} \right)
\\ c \left( \frac{d}{dr} + \frac{\kappa}{r} \right) & V(r) - 2c^2 \end{array} \right),
\eeqn{r.2}
where $c = 1/\alpha$. The quantum number $\kappa$ summarizes the orbital and total angular momentum quantum numbers $l$ and $j$ according to $j = |\kappa| - \dem$ and $l = j + \dem\, \mathrm{sgn}\, \kappa$. 

The large and small radial functions, $P_{n\kappa}$ and $Q_{n\kappa}$, are normalized according to the condition 
$\int_0^\infty \left\{ [P_{n\kappa}(r)]^2 + [Q_{n\kappa}(r)]^2 \right\} dr = 1$. 
At the origin \cite{Gr07}, they behave as 
\beq
P_{n\kappa}(r),\ Q_{n\kappa}(r) \arrow{r}{0} r^\gamma.
\eeqn{r.10}
The parameter $\gamma$ is defined by 
\beq
\gamma = \sqrt{\kappa^2 - (V_0/c)^2},
\eeqn{r.11}
where $V_0 = - \lim_{r \rightarrow 0} r V(r)$ is positive or null. 
The Dirac spinors are singular at the origin for $\gamma < 1$. 
This singularity can be important for hydrogenic ions with high nuclear charges $Z$. 

For a system described with the Dirac equation, the multipolar polarizability of a state $n\kappa m$ is given by 
\beq
\alpha^{(n\kappa m)}_{\lambda\mu} = (2j+1) \sum_{\kappa'} 
\left( \begin{array}{ccc} j' & \lambda & j \\  -m-\mu & \mu & m \end{array} \right)^2 
\alpha^{(n\kappa\kappa')}_{\lambda}. \eol
\eeqn{r.4}
The reduced polarizabilities read \cite{TZZ12} 
\beq
& & \alpha^{(n\kappa\kappa')}_{\lambda} = 2(2j'+1) 
\left( \begin{array}{ccc} j' & \lambda & j \\ -1/2 & 0 & 1/2 \end{array} \right)^2 \eol
& & \times \sum_{n'} \frac{\{\int_0^\infty [P_{n'\kappa'}(r) P_{n\kappa}(r) 
+ Q_{n'\kappa'}(r) Q_{n\kappa}(r)] r^\lambda dr\}^2}{E_{n'\kappa'}-E_{n\kappa}}, \eol
\eeqn{r.5}
where $P_{n\kappa}(r)$, $Q_{n\kappa}(r)$ and $P_{n'\kappa'}(r)$, $Q_{n'\kappa'}(r)$ are solutions of \rref{r.1} with respective energies $E_{n\kappa}$ and $E_{n'\kappa'}$. 
Like in the nonrelativistic case, the sum over $n'$ represents a sum over the discrete states and an integral over the continuum. 
Here, however, the continuum also involves negative energies. 
Degenerate states, i.e.\ states with $n' = n$ and $|\kappa'| = |\kappa|$, must be excluded 
in the hydrogenic case. 
Moreover, in this case,  we also exclude almost degenerate states with $n' = n$ but $|\kappa'| \neq |\kappa|$ 
because their small fine-structure energy differences are significantly affected by the Lamb shift 
and require a separate treatment \cite{Ya03,JH08} (see Sec.~\ref{sec:Coulomb}). 
Expression \rref{r.5} then tends to the nonrelativistic polarizabilities \rref{nr.3} when $c \rightarrow \infty$. 
The average or scalar polarizabilities are defined by 
\beq
\alpha^{(n\kappa)}_{\lambda} = \frac{1}{2j+1} \sum_{m=-j}^j \alpha^{(n\kappa m)}_{\lambda\mu} 
= \frac{1}{2\lambda+1} \sum_{\kappa'} \alpha^{(n\kappa\kappa')}_{\lambda}. \eol
\eeqn{r.6}

Here also a variant is possible. 
The inhomogeneous equation corresponding to \Eq{nr.4} reads 
\beq
(H_{\kappa'} - E_{n\kappa}) 
\left( \begin{array}{c} P^{(1)}_{n\kappa\kappa'}(r) \\ Q^{(1)}_{n\kappa\kappa'}(r) \end{array} \right) 
=  (1 - \cP_{n\kappa'}) r^\lambda \left( \begin{array}{c} P_{n\kappa}(r) \\ 
Q_{n\kappa}(r) \end{array} \right) \eol
\eeqn{r.7}
where $H_{\kappa'}$ is defined by \rref{r.2} with $\kappa'$ replacing $\kappa$ and $\cP_{n\kappa'}$ is the projector on
a state $n\kappa'$ degenerate or almost degenerate with the $n\kappa$ state, if any. 
Let $\widetilde{P}^{(1)}_{n\kappa\kappa'}$ and $\widetilde{Q}^{(1)}_{n\kappa\kappa'}$ be the radial components of the spinor obtained, if necessary, 
after application of the projector $(1 - \cP_{n\kappa'})$ on $(P^{(1)}_{n\kappa\kappa'}, Q^{(1)}_{n\kappa\kappa'})^T$, where $T$ means transposition. 
The reduced polarizabilities are given by the compact expressions 
\beq
& & \alpha^{(n\kappa\kappa')}_{\lambda} = 2(2j'+1) 
\left ( \begin{array}{ccc} j' & \lambda & j \\ -1/2 & 0 & 1/2 \end{array} \right )^2 \eol
& & \times \int_0^\infty [\widetilde{P}^{(1)}_{n\kappa\kappa'}(r) P_{n\kappa}(r) 
+ \widetilde{Q}^{(1)}_{n\kappa\kappa'}(r) Q_{n\kappa}(r)] r^\lambda dr.
\eeqn{r.8}


\subsection{Coulomb case}

In the Coulomb case, the potential is $V(r) = -Z/r$ in atomic units. 
Constant $V_0$ is equal to $Z \alpha c$. 
The energy of level $n\kappa$ is
\beq
E_{n \kappa} = - \dfrac{Z^2}{\cN (\cN + n - |\kappa| + \gamma)},
\eeqn{c.1}
with
\beq
\cN = [(n - |\kappa| + \gamma)^2 + (\alpha Z)^2]^{1/2}. 
\eeqn{c.2}
Analytically solving \Eq{r.7} is quite complicated, especially when projector $\cP_{n\kappa'}$ is present, 
but we can easily use it to make an optimal choice of pseudostates for an approximation of \Eq{r.5}. 
Two cases must be considered. 

When $|\kappa'| = |\kappa|$, $P^{(1)}_{n\kappa\kappa'}$ and $Q^{(1)}_{n\kappa\kappa'}$ have the same behavior at the origin as in \Eq{r.10} because parameter $\gamma$ given by \Eq{r.11} is the same. 
Hence the solution of \Eq{r.7} is given by $r^\gamma \exp(-Zr/\cN)$ multiplied by polynomials. 
The pseudostates in \Eq{r.5} should contain the same $r^\gamma$ factor and the same exponential. 

When $|\kappa'| \neq |\kappa|$, however, there is no simple analytical solution of \Eq{r.7} because $\gamma'$ corresponding to $\kappa'$ differs from $\gamma$. 
This is exemplified by the exact ground-state polarizabilities which are the sum of a simple $\kappa'=1$ term and a complicated $\kappa'=-2$ term \cite{SM04}. 


\section{Lagrange-mesh method}
\label{sec:LMM}

\subsection{Mesh equations}

The principles of the Lagrange-mesh method are described in Refs.~\cite{BH86,VMB93,Ba06} and its application to the Dirac equation is presented in \Ref{BFG14}. 
The mesh points $x_j$ are defined by \cite{BH86} 
\beq
L_N^{\alpha}(x_j) = 0,
\eeqn{Lag.1}
where $j = 1$ to $N$ and $L_N^{\alpha}$ is a generalized Laguerre polynomial depending on parameter $\alpha$ \cite{AS65}. 
This mesh is associated with a Gauss quadrature 
\beq
\int_0^\infty g(x) \, dx \approx \sum^N_{k=1} \lambda_k g(x_k), 
\eeqn{Lag.2}
with the weights $\lambda_k$. 
The Gauss quadrature is exact for the Laguerre weight function $x^{\alpha} e^{-x}$ multiplied by any polynomial of degree at most $2N-1$ \cite{Sz67}. 

The regularized Lagrange functions are defined by \cite{Ba95,BHV02,Ba06} 
\beq
\hat{f}_j^{(\alpha)} (x) & = & (-1)^j \sqrt{\frac{N !}{\Gamma (N + \alpha + 1) x_j}} \, \frac{L_N^{\alpha}(x)}{x-x_j} \, x^{\alpha/2+1} e^{-x/2}. \eol
\eeqn{Lag.3}
The functions $\hat{f}_j^{(\alpha)}(x)$ are polynomials of degree $N-1$ multiplied by $x$ and by the square root of the Laguerre weight $x^{\alpha} \exp(-x)$. The Lagrange functions satisfy the Lagrange conditions 
\beq
\hat{f}_j^{(\alpha)}(x_i) = \lambda_i^{-1/2} \delta_{ij}.
\eeqn{Lag.4}
They are orthonormal at the Gauss-quadrature approximation. Condition \rref{Lag.4} drastically simplifies the expressions calculated with the Gauss quadrature. 

The radial functions $P_{n\kappa}(r)$ and $Q_{n\kappa}(r)$ are expanded in regularized Lagrange functions \rref{Lag.3} as 
\beq
P_{n\kappa}(r) = h^{-1/2} \sum_{j=1}^{N} \; p_{n\kappa j} \hat{f}_j^{(\alpha)}(r/h),
\eoln{Lag.5}
Q_{n\kappa}(r) = h^{-1/2} \sum_{j=1}^{N} \; q_{n\kappa j} \hat{f}_j^{(\alpha)}(r/h),
\eeqn{Lag.6}
where $h$ is a scaling parameter aimed at adapting the mesh points $hx_i$ to the physical extension of the problem 
and $\sum_{j=1}^N \left( p_{n\kappa j}^2 + q_{n\kappa j}^2 \right) = 1$ ensures the normalization of $P_{n\kappa}$ and $Q_{n\kappa}$. 

The parameter $\alpha = 2 ( \gamma - 1)$ can be selected so that the Lagrange functions behave as $r^\gamma$ near the origin \cite{BFG14}. 
Here, another choice $\alpha = 2 ( \gamma - |\kappa|)$ is preferable as explained below. 
The basis functions then behave as $r^{\gamma-|\kappa|+1}$ but the physical $r^\gamma$ behavior can be simulated by linear combinations. 
In the Coulomb case, the correct exponential behavior of the components is obtained with $h = \cN/2Z$. 
Expansions with $N = n + |\kappa|$ such functions are able to exactly reproduce the large and small hydrogenic components. 
Moreover matrix elements of $r^\lambda$ with $\lambda \ge -2$ are exactly obtained with $2N - 1 \ge 2n + 2|\kappa| + \lambda$ or $N > n + |\kappa| + \dem \lambda$ \cite{BHV02,Ba06}. 

Let us introduce expansions \rref{Lag.5} and \rref{Lag.6} in the coupled radial Dirac equations \rref{r.1}. 
Projecting on the Lagrange functions and using the associated Gauss quadrature leads to the $2N \times 2N$ Hamiltonian matrix 
\beq
\ve{H}_\kappa = \left( \begin{array}{cc} 
V(hx_i) \delta_{ij} & 
\frac{c}{h} \left( D_{ji}^G + \frac{\kappa}{x_i} \delta_{ij} \right) \\
\frac{c}{h} \left( D_{ij}^G + \frac{\kappa}{x_i} \delta_{ij} \right) &
(V(hx_i) -2c^2) \delta_{ij} 
\end{array} \right)
\eeqn{Lag.7}
with a $2 \times 2$ block structure, where 
\beq
D_{i \neq j}^G = (-1)^{i-j} \sqrt{\frac{x_i}{x_j}}\, \frac{1}{x_i-x_j},
\quad
D_{ii}^G = \frac{1}{2x_i}.
\eeqn{Lag.8}
Expressions \rref{Lag.8} are the matrix elements $\la \hat{f}_i^{(\alpha)} |d/dx| \hat{f}_j^{(\alpha)} \ra$ calculated at the Gauss-quadrature approximation. 
This corresponds to choosing the Gauss quadrature named `Gauss(2,1)' in \Ref{BFG14}. 

In the Coulomb case, if $N \ge n + |\kappa|$ and $h = \cN/2Z$, 
one of the eigenvalues of $\ve{H}_\kappa$ is the exact energy $E_{n\kappa}$ and the corresponding eigenvector $(p_{n\kappa 1}, p_{n\kappa 2}, \dots, p_{n\kappa N}, q_{n\kappa 1}, q_{n\kappa 2}, \dots, q_{n\kappa N})^T$ provides the coefficients of the exact eigenfunctions in the expansions \rref{Lag.5} and \rref{Lag.6} \cite{BFG14}. 
For other potentials, if $N$ is large enough and $h$ well chosen, some negative energies above $-2c^2$ correspond to physical energies. 
The corresponding eigenvectors provide approximations of the wave functions. 


\subsection{Polarizabilities on a Lagrange mesh}

As proven in Appendix \ref{sec:A}, the Lagrange-mesh approximations of Eqs.~\rref{r.5} and \rref{r.8} are identical if the same scaling parameter $h$ is used. 
However, the calculation is simpler with \Eq{r.5}, specially when degenerate levels must be eliminated. 

Let $E^{(k)}_{\kappa'}$, $k = 1, \dots, 2N$, be the eigenvalues of matrix $\ve{H}_{\kappa'}$ defined in \Eq{Lag.7} with $\kappa'$ replacing $\kappa$. The corresponding eigenvectors contain the coefficients $p^{(k)}_{\kappa' j}$ and $q^{(k)}_{\kappa' j}$ of the components $P^{(k)}_{\kappa'}$ and $Q^{(k)}_{\kappa'}$ of the pseudostates. 
These pseudostates are obtained with the same $h$ value as for the $n\kappa$ state. 
They have no physical meaning. 

Approximate reduced polarizabilities can be obtained from \Eq{r.5} as 
\beq
& & \alpha^{(n\kappa\kappa')}_{\lambda} = 2(2j'+1) 
\left( \begin{array}{ccc} j' & \lambda & j \\ -1/2 & 0 & 1/2 \end{array} \right)^2 \eol
& & \times \sum_{k=1}^N \frac{\{\int_0^\infty [P^{(k)}_{\kappa'}(r) P_{n\kappa}(r) 
+ Q^{(k)}_{\kappa'}(r) Q_{n\kappa}(r)] r^\lambda dr\}^2}{E^{(k)}_{\kappa'}-E_{n\kappa}}. \eol
\eeqn{Lag.10}
The integral in \rref{Lag.10} is calculated with the Gauss quadrature as 
\beq
& & \int_0^\infty [P^{(k)}_{\kappa'}(r) P_{n\kappa}(r) + Q^{(k)}_{\kappa'}(r) Q_{n\kappa}(r)] r^\lambda dr \eol
& & \approx h^{\lambda} \sum_{j=1}^N 
[p^{(k)}_{\kappa' j} p_{n\kappa j} + q^{(k)}_{\kappa' j} q_{n\kappa j}] x_j^\lambda.
\eeqn{Lag.11}
Contrary to the nonrelativistic case, the partial polarizabilities of the hydrogenic atoms are not always exactly given by the Lagrange-mesh method. 
For $|\kappa'| = |\kappa|$, they are exact for $N > n + |\kappa| + \dem \lambda$, 
but they are not exact for $|\kappa'| \neq |\kappa|$. 
Indeed, the eigenfunctions of $H_{\kappa'}$ behave at the origin as $r^{\gamma'}$ 
where $\gamma' = \sqrt{\kappa^{\prime 2} - (V_0/c)^2} \neq \gamma$. 
The solution of \Eq{r.7} does not involve any more a simple polynomial. 
In this case, we nevertheless choose to keep Eqs.~\rref{Lag.10} and \rref{Lag.11} as an approximation. 
As shown in section \ref{sec:Coulomb}, when $N$ is large enough, 
accurate values of the polarizabilities can be obtained for low $Z$ values since $\gamma'$ is then close to $\gamma$. 


\subsection{Polarizabilities with three meshes}

When $|\kappa'| \neq |\kappa|$, a better approximation than \Eq{Lag.11} is possible. 
Let us calculate the pseudostates with $\alpha' = 2 ( \gamma' - |\kappa'|)$ in place of $\alpha = 2 ( \gamma - |\kappa|)$, i.e.\ matrix $\ve{H}_{\kappa'}$ is calculated on a different mesh $hx'_j$. 
Hence the pseudostates have the exact behavior $r^{\gamma'}$ at the origin. 
Notice that the values of $\alpha$ and $\alpha'$ are close to each other, much closer than with the choices $\alpha' = 2 ( \gamma' - 1)$ and $\alpha = 2 ( \gamma - 1)$. 
The integrand in \Eq{Lag.11} explicitly contains $r^{\gamma+\gamma'}$. 
In the Coulomb case with $h = \mathcal{N}/2Z$, it is even the product of $r^{\gamma+\gamma'} \exp(-2Zr/\cN)$ and a polynomial. 
An accurate calculation with a Gauss quadrature is possible in \Eq{Lag.10} by choosing a third mesh $h \bar{x}_i$ where the $\bar{x}_i$ correspond to the weight function $x^{\bar{\alpha}} \exp(-x)$ with the average value 
\beq
\bar{\alpha} = \dem (\alpha + \alpha').
\eeqn{Lag.12}
The corresponding weights are denoted as $\bar{\lambda}_i$. 
For simplicity, we keep the same number $N$ of mesh points for the three meshes but this is not at all mandatory. 

Approximation \rref{Lag.11} is replaced by the expression 
\beq
& & \int_0^\infty [P^{(k)}_{\kappa'}(r) P_{n\kappa}(r) 
+ Q^{(k)}_{\kappa'}(r) Q_{n\kappa}(r)] r^\lambda dr \eol
& & \approx h^{\lambda} \sum_{j,j'=1}^N 
[p^{(k)}_{\kappa' j'} p_{n\kappa j} + q^{(k)}_{\kappa' j'} q_{n\kappa j}] 
I^\lambda_{j'j}
\eeqn{Lag.13}
where 
\beq
I^\lambda_{j'j} & = & \int_0^\infty \hat{f}_{j'}^{(\alpha')}(x) 
x^\lambda \hat{f}_j^{(\alpha)}(x) dx \eol
& \approx & \sum_{i=1}^N \bar{\lambda}_{i} \hat{f}_{j'}^{(\alpha')}(\bar{x}_{i}) 
\bar{x}_{i}^\lambda \hat{f}_j^{(\alpha)}(\bar{x}_{i}).
\eeqn{Lag.14}
Evaluating integral \rref{Lag.14} requires the explicit computation of Lagrange functions. 
Some remarks on their numerical calculation can be found in Appendix \ref{sec:B}. 

The Gauss quadrature in \Eq{Lag.13} is exact in the Coulomb case if $2N-1 \geq N + n + |\kappa| + \lambda$ or $N \geq n + |\kappa| + \lambda + 1$, 
but the reduced polarizability \rref{Lag.10} is not exact because the corresponding solution of \Eq{r.7} is an approximation. 


\section{Numerical results}
\label{sec:res}
\subsection{Hydrogenic atoms}
\label{sec:Coulomb}

We first calculate ground-state polarizabilities for the Dirac-Coulomb problem, where $V(r) = -Z/r$ in atomic units. 
We consider static dipole to hexadecapole polarizabilities of hydrogenic ions for $Z$ values comprised between 1 and 100. 

\begin{figure}[!ht]
\begin{center}
\includegraphics[scale=0.45]{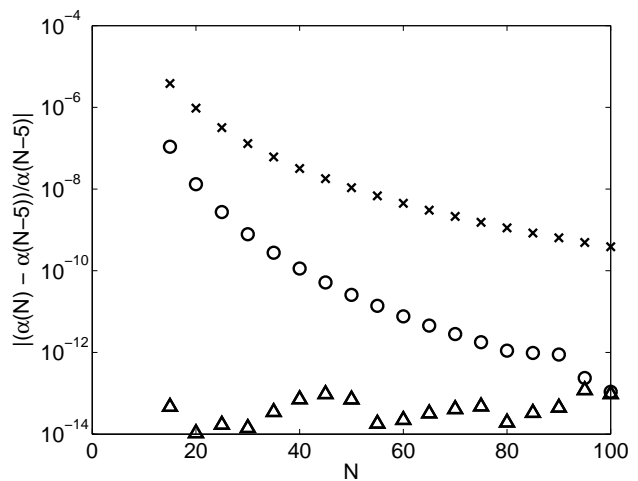}
\end{center}
\caption{Convergence of the $1s_{1/2}$ partial static dipole polarizabilities of the $Z=100$ hydrogenic ion 
as a function of the number $N$ of mesh points. 
Contributions from the $p_{1/2}$ ($\kappa = +1$) states (triangles) and the $p_{3/2}$ ($\kappa = -2$) states (crosses for one mesh, circles for three meshes).}
\label{fig_diff_1s12_partiel_Z100}
\end{figure}

Before giving numerical values for the $1s_{1/2}$ ground-state static dipole polarizabilities, 
we first compare the single-mesh and three-mesh methods presented in the previous section. 
In \figurename{~\ref{fig_diff_1s12_partiel_Z100}}, the convergence of the two partial polarizabilities as a function of the number of mesh points $N$ is evaluated by the relative difference $\vert \alpha(N) - \alpha(N-5) \vert / \vert \alpha(N-5) \vert$, 
where $N=15$ to $100$ by steps of $5$. 
The case $Z=100$ is chosen in order to emphasize the effect of using three different meshes instead of a single mesh, as discussed in Sec.~\ref{sec:LMM}.

In this figure, crosses correspond to the contribution of the $p_{3/2}$ states ($\kappa = -2$) computed with one mesh, 
while circles correspond to the same contribution but computed with three meshes. 
Triangles correspond to the contribution of the $p_{1/2}$ states ($\kappa = +1$) computed indifferently with one mesh or three meshes. 
Indeed, the results of both approaches are exactly the same for this contribution, for which $\vert \kappa' \vert = \vert \kappa \vert$. 
Then $\bar{\alpha}=\alpha'=\alpha$ in \Eq{Lag.12} and the three meshes are degenerate into a single one. 
It is thus no longer required to explicitly compute the Lagrange functions, and both computations use Eqs.~\rref{Lag.10} and \rref{Lag.11}. 
The calculation of the partial polarizability is exactly given by the Lagrange-mesh method. 
The polarizabilities represented by triangles are numerically exact

\pagebreak

\onecolumngrid

\begin{table}[!ht]
\caption{Static $\lambda$-multipole polarizabilities (in a.u.) of the $1s_{1/2}$ ground state of hydrogenic ions. 
Comparison between $N$ and $N+2$ mesh points and with benchmark values \cite{TZZ12}. 
Powers of 10 are indicated within brackets.}
\label{table_1s12_Z1_100}
\begin{center}
\begin{tabular}{c r c r c l c r c l c l}
\hline
\hline
\vspace{0.1cm}
          &     & \hspace{0.5cm} & \multicolumn{7}{c}{Lagrange-mesh}                                                       & \hspace{0.5cm} & \Ref{TZZ12}                               \\
\cline{4-10}
\vspace{0.1cm}
$\lambda$ & $Z$ & \hspace{0.5cm} & $N$ & & $\alpha_{\lambda}^{(1s_{1/2})}$   & & $N$ & & $\alpha_{\lambda}^{(1s_{1/2})}$   & \hspace{0.5cm} & $\alpha_{\lambda}^{(1s_{1/2})}$ ($N=400$) \\
\hline
1         & 1   & \hspace{0.5cm} & 6   & & $4.499\,751\,495\,177\,656$       & & 8   & & $4.499\,751\,495\,177\,639$       & \hspace{0.5cm} & $4.499\,751\,495\,177\,639\,267\,4$       \\
          & 2   & \hspace{0.5cm} & 8   & & $0.281\,187\,874\,918\,502$       & & 10  & & $0.281\,187\,874\,918\,506$       & \hspace{0.5cm} & $0.281\,187\,874\,918\,503\,235\,4$       \\
          & 20  & \hspace{0.5cm} & 20  & & $2.750\,523\,499\,061\,9\,[-5]$   & & 22  & & $2.750\,523\,499\,064\,3\,[-5]$   & \hspace{0.5cm} & $2.750\,523\,499\,062\,579\,08[-5]$       \\
          & 40  & \hspace{0.5cm} & 40  & & $1.604\,002\,839\,548\,4\,[-6]$   & & 42  & & $1.604\,002\,839\,548\,7\,[-6]$   & \hspace{0.5cm} & $1.604\,002\,839\,548\,263\,7[-6]$        \\
          & 60  & \hspace{0.5cm} & 50  & & $2.797\,090\,474\,417\,6\,[-7]$   & & 52  & & $2.797\,090\,474\,417\,0\,[-7]$   & \hspace{0.5cm} & $2.797\,090\,474\,417\,353[-7]$           \\
          & 80  & \hspace{0.5cm} & 70  & & $7.256\,230\,363\,582\,9\,[-8]$   & & 72  & & $7.256\,230\,363\,582\,7\,[-8]$   & \hspace{0.5cm} & $7.256\,230\,363\,582\,21[-8]$            \\
\vspace{0.2cm}  
          & 100 & \hspace{0.5cm} & 100 & & $2.168\,647\,587\,492\,2\,[-8]$   & & 102 & & $2.168\,647\,587\,492\,9\,[-8]$   & \hspace{0.5cm} & $2.168\,647\,587\,493\,68[-8]$            \\
2         & 1   & \hspace{0.5cm} & 8   & & $14.998\,829\,822\,856\,73$       & & 10  & & $14.998\,829\,822\,856\,48$       & \hspace{0.5cm} & $14.998\,829\,822\,856\,441\,699$         \\
          & 2   & \hspace{0.5cm} & 8   & & $0.234\,301\,867\,935\,799$       & & 10  & & $0.234\,301\,867\,935\,789$       & \hspace{0.5cm} & $0.234\,301\,867\,935\,791\,100$          \\
          & 20  & \hspace{0.5cm} & 20  & & $2.271\,146\,583\,055\,3\,[-7]$   & & 22  & & $2.271\,146\,583\,050\,7\,[-7]$   & \hspace{0.5cm} & $2.271\,146\,583\,050\,793[-7]$           \\
          & 40  & \hspace{0.5cm} & 40  & & $3.218\,326\,876\,369\,1\,[-9]$   & & 40  & & $3.218\,326\,876\,369\,6\,[-9]$   & \hspace{0.5cm} & $3.218\,326\,876\,369\,0[-9]$             \\
          & 60  & \hspace{0.5cm} & 50  & & $2.371\,147\,053\,044\,9\,[-10]$  & & 52  & & $2.371\,147\,053\,044\,6\,[-10]$  & \hspace{0.5cm} & $2.371\,147\,053\,044[-10]$               \\
          & 80  & \hspace{0.5cm} & 70  & & $3.196\,013\,748\,395\,1\,[-11]$  & & 72  & & $3.196\,013\,748\,395\,1\,[-11]$  & \hspace{0.5cm} & $3.196\,013\,748\,39[-11]$                \\
\vspace{0.2cm}  
          & 100 & \hspace{0.5cm} & 100 & & $5.405\,559\,183\,469\,5\,[-12]$  & & 102 & & $5.405\,559\,183\,470\,7\,[-12]$  & \hspace{0.5cm} & $5.405\,559\,183\,5[-12]$                 \\
3         & 1   & \hspace{0.5cm} & 8   & & $131.237\,821\,447\,843\,0$       & & 10  & & $131.237\,821\,447\,846\,0$       & \hspace{0.5cm} & $131.237\,821\,447\,844\,662$             \\
          & 2   & \hspace{0.5cm} & 8   & & $0.512\,505\,037\,523\,772$       & & 10  & & $0.512\,505\,037\,523\,776$       & \hspace{0.5cm} & $0.512\,505\,037\,523\,770\,47$           \\
          & 20  & \hspace{0.5cm} & 20  & & $4.938\,640\,072\,266\,9\,[-9]$   & & 22  & & $4.938\,640\,072\,274\,8\,[-9]$   & \hspace{0.5cm} & $4.938\,640\,072\,269\,2[-9]$             \\
          & 40  & \hspace{0.5cm} & 40  & & $1.717\,671\,116\,720\,7\,[-11]$  & & 42  & & $1.717\,671\,116\,720\,8\,[-11]$  & \hspace{0.5cm} & $1.717\,671\,116\,72[-11]$                \\
          & 60  & \hspace{0.5cm} & 50  & & $5.443\,579\,080\,006\,2\,[-13]$  & & 52  & & $5.443\,579\,080\,006\,2\,[-13]$  & \hspace{0.5cm} & $5.443\,579\,080[-13]$                    \\
          & 80  & \hspace{0.5cm} & 70  & & $3.921\,694\,887\,294\,4\,[-14]$  & & 72  & & $3.921\,694\,887\,294\,5\,[-14]$  & \hspace{0.5cm} & $3.921\,694\,89[-14]$                     \\
\vspace{0.2cm}  
          & 100 & \hspace{0.5cm} & 100 & & $3.923\,335\,154\,080\,2\,[-15]$  & & 102 & & $3.923\,335\,154\,081\,2\,[-15]$  & \hspace{0.5cm} & $3.923\,335\,2[-15]$                      \\ 
4         & 1   & \hspace{0.5cm} & 8   & & $2\,126.028\,674\,498\,991$       & & 10  & & $2\,126.028\,674\,499\,147$       & \hspace{0.5cm} & $2\,126.028\,674\,499\,128\,83$           \\
          & 2   & \hspace{0.5cm} & 8   & & $2.075\,551\,546\,061\,163$       & & 10  & & $2.075\,551\,546\,061\,205$       & \hspace{0.5cm} & $2.075\,551\,546\,061\,205\,19$           \\
          & 20  & \hspace{0.5cm} & 20  & & $1.991\,062\,443\,016\,7\,[-10]$  & & 22  & & $1.991\,062\,443\,018\,8\,[-10]$  & \hspace{0.5cm} & $1.991\,062\,443\,017[-10]$               \\
          & 40  & \hspace{0.5cm} & 40  & & $1.707\,067\,336\,464\,2\,[-13]$  & & 42  & & $1.707\,067\,336\,464\,5\,[-13]$  & \hspace{0.5cm} & $1.707\,067\,337[-13]$                    \\
          & 60  & \hspace{0.5cm} & 50  & & $2.345\,208\,224\,082\,4\,[-15]$  & & 52  & & $2.345\,208\,224\,082\,5\,[-15]$  & \hspace{0.5cm} & $2.345\,208\,2[-15]$                      \\
          & 80  & \hspace{0.5cm} & 70  & & $9.141\,669\,892\,326\,1\,[-17]$  & & 72  & & $9.141\,669\,892\,323\,9\,[-17]$  & \hspace{0.5cm} & $9.141\,67[-17]$                          \\
\vspace{0.1cm}
          & 100 & \hspace{0.5cm} & 100 & & $5.514\,202\,246\,345\,4\,[-18]$  & & 102 & & $5.514\,202\,246\,347\,0\,[-18]$  & \hspace{0.5cm} & $5.514\,2[-18]$                           \\
\hline
\hline
\end{tabular}
\end{center}
\end{table}

\twocolumngrid

\noindent for all $N$ values, and oscillate below $10^{-13}$. 
This is not the case for the contribution from the $p_{3/2}$ states, for which $\vert \kappa' \vert \neq \vert \kappa \vert$. 
For $Z=100$, the results of both approaches are poor for low $N$ values but the errors progressively decrease when $N$ increases. 
The results with three meshes are significantly better than the ones with a single mesh, as shown in \figurename{~\ref{fig_diff_1s12_partiel_Z100}}.  Indeed, the circles are from one to three orders of magnitude lower than the crosses, reaching the value of $10^{-13}$ for $N=100$, 
while the result with a single mesh only reaches $4 \times 10^{-9}$ for the same number of mesh points. 
In the following, all the results displayed in the tables are obtained from computations performed with three meshes.

\tablename{~\ref{table_1s12_Z1_100}} displays the values of static multipole polarizabilities for the $1s_{1/2}$ ground state of hydrogenic ions, with $Z$ values between $1$ to $100$ and given values of $N$. 
The optimal scaling parameter $h$ is used, i.e. $h=\mathcal{N}/2Z$, where $\mathcal{N}$ is given by \Eq{c.2}. 
Four different multipoles are considered, from $\lambda=1$ to $4$. 
The significant digits of the results can be estimated by a comparison with $N+2$ mesh points. 
In \Ref{TZZ12}, Tang \textit{et al.} used  the Galerkin method to provide very accurate numerical values with a basis of $400$ B-splines. 
These results will be used as benchmark values, in order to test the precision of the Lagrange-mesh method. 

One observes a relative difference better than $10^{-12}$ between Lagrange-mesh computations and these benchmark values, some results agreeing with up to $15$ figures. 
An important fact to mention is the use of significantly fewer mesh points in the Lagrange-mesh computation than in the B-spline Galerkin method, as reported in the table. 
Indeed, only $6$ mesh points are sufficient to obtain a $10^{-14}$ accuracy with $\lambda=Z=1$. 
For high $Z$, up to $Z=100$, the number of mesh points does not go beyond $100$. 
Moreover, for values from quadrupole to hexadecapole, the Lagrange-mesh computations provide more significant digits than the B-spline Galerkin method when $Z$ increases. 
Regarding the stability of the Lagrange-mesh results, one can observe that at most the last two digits are varying from $N$ to $N+2$ mesh points.

\begin{figure}[!ht]
\begin{center}
\includegraphics[scale=0.45]{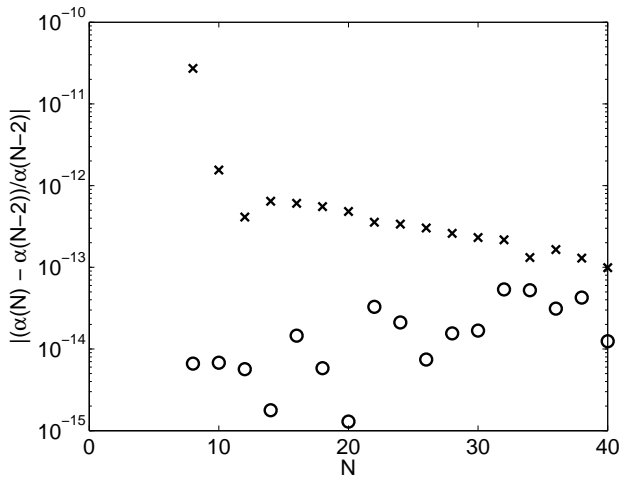}
\end{center}
\caption{Convergence of the $2p_{3/2}$ static dipole polarizability of the hydrogen atom as a function of the number $N$ of mesh points. 
Use of one mesh (crosses) and three meshes (circles).}
\label{fig_diff_2p32_Z1}
\end{figure}

Let us now consider relativistic polarizabilities of excited states of hydrogenic ions, for which no values exist in the literature. 
In this paper, we focus on the $n=2$ states, although other bound states can be treated as easily and accurately with this method. 
\figurename{~\ref{fig_diff_2p32_Z1}} shows the convergence of the $2p_{3/2}$ static dipole polarizability of the hydrogen atom as a function of the number of mesh points $N$, evaluated by the relative difference $\vert \alpha(N) - \alpha(N-2) \vert / \vert \alpha(N-2) \vert$, 
where $N=8$ to $40$ by steps of $2$. 
The case $Z=1$ is sufficient to emphasize the improvement brought by using three different meshes instead of a single one. 
Crosses and circles respectively correspond to the use of one mesh and three meshes. 
One clearly observes the effect of using three meshes on the accuracy of the results. 
Indeed, all the errors represented by circles progressively increase with $N$ because of larger rounding errors, 
but stay below $10^{-13}$ while the errors represented by crosses decrease from $3 \times 10^{-11}$ when $N=8$ to $10^{-13}$ when $N=40$. 
The total static dipole polarizability $\alpha_{1}^{(2p_{3/2})}$ has three contributions: from the $d_{3/2}$ states for which $\vert \kappa' \vert = \vert \kappa \vert$, and from the $s_{1/2}$ and $d_{5/2}$ states for which $\vert \kappa' \vert \neq \vert \kappa \vert$. 
For these last two contributions, the results obtained with three meshes are thus significantly better than the ones with a single mesh, 
while for the first contribution both approaches give exactly the same results, as discussed above for \figurename{~\ref{fig_diff_1s12_partiel_Z100}}.

\tablename{~\ref{table_n2_Z1-100}} presents the values of the static dipole and quadrupole polarizabilities for the $n=2$ states of hydrogenic ions, with $Z$ values between $1$ to $100$ and given values of $N$. 
The optimal scaling parameter $h$ is used, i.e. $h=\mathcal{N}/2Z$. 
The significant digits of the results are estimated by a comparison with $N+2$ mesh points. 
For each state and each $Z$ value, one observes an accuracy of at least $10^{-13}$.
The results for the hydrogen atom are obtained with only $6$ mesh points for $\lambda=1$ and $8$ mesh points for $\lambda=2$. 
For low $Z$ values, the dipole polarizabilities are very close to the exact nonrelativistic ones \cite{KMM97,Ba12}. 
Indeed, the relativistic effects are very weak. 
When the fine-structure constant tends to zero, the $2s_{1/2}$ polarizability tends to $120 Z^{-4}$ a.u.\ and 
the $2p_{1/2}$ and $2p_{3/2}$ states degenerate into a single $2p$ state whose nonrelativistic polarizability is $176 Z^{-4}$ a.u.  
When $Z$ increases, the relativistic effects become more important than for $Z=1$ and the polarizabilities are no longer close to the exact nonrelativistic values.

The limited Taylor expansion of the $2s$ polarizability 
\beq
\alpha_{1}^{(2s_{1/2})} = \frac{120}{Z^4} \left[ 1 - \frac{367}{240} (\alpha Z)^2 + 0.575887 (\alpha Z)^4 \right] \eol
\eeqn{hyd:eq0}
given by Eq.~(19) of \Ref{Ya03} is in perfect agreement with the values in Table \ref{table_n2_Z1-100} for $Z = 1$ and 2. 
For $Z = 10$, its relative accuracy is still better than $10^{-9}$. 
For $Z = 50$ and 100, it drops to $10^{-5}$ and $4 \times 10^{-4}$, respectively, 
still a fair approximation despite that $\alpha Z$ progressively approaches unity. 

As explained in Sec.~\ref{sec:Dirac}, in the hydrogenic case, degenerate states ($n'=n$ and $\vert \kappa' \vert = \vert \kappa \vert$) 
and almost degenerate states ($n'=n$ but $\vert \kappa' \vert \neq \vert \kappa \vert$) are excluded from the polarizability of a $n\kappa$ state. 
All values displayed up to now do not take these states into account. 
However, in reality, these states are not exactly degenerate and their effect must be included in the polarizability. 
Their contributions to the total static polarizability $\bar{\alpha}^{(n\kappa)}$ of a $n\kappa$ state can be computed with the reduced matrix elements appearing in the numerator of \Eq{r.5} and with exact values for the differences of energies appearing in the denominator of this expression. 
Let us illustrate this consideration with the example of the $n=2$ states of hydrogenic ions. 
The total static dipole polarizabilities read
\beq
\bar{\alpha}_{1}^{(2s_{1/2})} & = & \alpha_{1}^{(2s_{1/2})} + \dfrac{F(2p_{1/2},2s_{1/2})}{E_{2p_{1/2}} - E_{2s_{1/2}}} + \dfrac{F(2p_{3/2},2s_{1/2})}{E_{2p_{3/2}} - E_{2s_{1/2}}}, \eol
\eoln{hyd:eq1a}
\bar{\alpha}_{1}^{(2p_{1/2})} & = & \alpha_{1}^{(2p_{1/2})} + \dfrac{F(2s_{1/2},2p_{1/2})}{E_{2s_{1/2}} - E_{2p_{1/2}}}, 
\eoln{hyd:eq1b}
\bar{\alpha}_{1}^{(2p_{3/2})} & = & \alpha_{1}^{(2p_{3/2})} + \dfrac{F(2s_{1/2},2p_{3/2})}{E_{2s_{1/2}} - E_{2p_{3/2}}}
\eeqn{hyd:eq1c}
for each of the $n=2$ states, where the quantities $\alpha_{1}^{(2l_j)}$

\pagebreak

\onecolumngrid

\begin{table}[!ht]
\caption{Static $\lambda$-multipole polarizabilities (in a.u.) of the $n=2$ states of hydrogenic ions. 
Significants digits are estimated by comparison with $N+2$ mesh points. 
Powers of 10 are indicated within brackets.}
\label{table_n2_Z1-100}
\begin{center}
\begin{tabular}{c r r c l c l c l}
\hline
\hline
\vspace{0.1cm}
$\lambda$ & $Z$ & $N$ & \hspace{0.5cm} & $\alpha_{\lambda}^{(2s_{1/2})}$ & \hspace{0.5cm} & $\alpha_{\lambda}^{(2p_{1/2})}$ & \hspace{0.5cm} & $\alpha_{\lambda}^{(2p_{3/2})}$ \\
\hline
1         & 1   & 6   & \hspace{0.5cm} & $119.990\,228\,572\,41$         & \hspace{0.5cm} & $175.987\,219\,883\,326$        & \hspace{0.5cm} & $175.997\,972\,047\,628$        \\
          & 2   & 8   & \hspace{0.5cm} & $7.497\,557\,290\,076\,1$       & \hspace{0.5cm} & $10.996\,805\,127\,139\,4$      & \hspace{0.5cm} & $10.999\,493\,039\,445\,4$      \\
          & 10  & 10  & \hspace{0.5cm} & $1.190\,247\,972\,335\,[-2]$    & \hspace{0.5cm} & $1.747\,240\,513\,459\,[-2]$    & \hspace{0.5cm} & $1.757\,975\,693\,736\,[-2]$    \\
          & 20  & 20  & \hspace{0.5cm} & $7.257\,668\,813\,803\,[-4]$    & \hspace{0.5cm} & $1.068\,257\,492\,237\,[-3]$    & \hspace{0.5cm} & $1.094\,967\,210\,025\,[-3]$    \\
          & 30  & 30  & \hspace{0.5cm} & $1.374\,866\,994\,958\,[-4]$    & \hspace{0.5cm} & $2.032\,917\,604\,564\,[-4]$    & \hspace{0.5cm} & $2.150\,684\,232\,965\,[-4]$    \\  
          & 40  & 40  & \hspace{0.5cm} & $4.096\,360\,848\,269\,[-5]$    & \hspace{0.5cm} & $6.097\,028\,078\,230\,[-5]$    & \hspace{0.5cm} & $6.752\,120\,183\,672\,[-5]$    \\
          & 50  & 40  & \hspace{0.5cm} & $1.548\,718\,165\,822\,[-5]$    & \hspace{0.5cm} & $2.325\,561\,392\,593\,[-5]$    & \hspace{0.5cm} & $2.738\,874\,396\,679\,[-5]$    \\  
          & 60  & 50  & \hspace{0.5cm} & $6.740\,709\,021\,419\,[-6]$    & \hspace{0.5cm} & $1.023\,760\,773\,010\,[-6]$    & \hspace{0.5cm} & $1.305\,852\,703\,883\,[-6]$    \\
          & 70  & 60  & \hspace{0.5cm} & $3.199\,464\,332\,281\,[-6]$    & \hspace{0.5cm} & $4.929\,154\,351\,302\,[-6]$    & \hspace{0.5cm} & $6.960\,240\,041\,733\,[-6]$    \\  
          & 80  & 70  & \hspace{0.5cm} & $1.598\,609\,255\,742\,[-6]$    & \hspace{0.5cm} & $2.506\,652\,478\,893\,[-6]$    & \hspace{0.5cm} & $4.026\,825\,704\,514\,[-6]$    \\
          & 90  & 80  & \hspace{0.5cm} & $8.183\,632\,847\,570\,[-7]$    & \hspace{0.5cm} & $1.310\,940\,334\,525\,[-6]$    & \hspace{0.5cm} & $2.483\,072\,557\,892\,[-6]$    \\ 
\vspace{0.2cm} 
          & 100 & 100 & \hspace{0.5cm} & $4.186\,326\,200\,652\,[-7]$    & \hspace{0.5cm} & $6.876\,679\,310\,957\,[-7]$    & \hspace{0.5cm} & $1.613\,980\,660\,494\,[-6]$    \\  
2         & 1   & 8   & \hspace{0.5cm} & $16\,318.452\,858\,385$         & \hspace{0.5cm} & $5\,183.438\,096\,590\,6$       & \hspace{0.5cm} & $5\,183.929\,162\,202\,5$       \\
          & 2   & 8   & \hspace{0.5cm} & $254.903\,312\,000\,07$         & \hspace{0.5cm} & $80.964\,884\,615\,008$         & \hspace{0.5cm} & $80.995\,572\,769\,725$         \\
          & 10  & 10  & \hspace{0.5cm} & $1.616\,572\,636\,637\,[-2]$    & \hspace{0.5cm} & $5.127\,998\,385\,845\,[-3]$    & \hspace{0.5cm} & $5.176\,923\,204\,048\,[-3]$    \\  
          & 20  & 20  & \hspace{0.5cm} & $2.454\,410\,134\,238\,[-4]$    & \hspace{0.5cm} & $7.753\,548\,824\,928\,[-5]$    & \hspace{0.5cm} & $8.055\,903\,187\,452\,[-5]$    \\
          & 30  & 30  & \hspace{0.5cm} & $2.052\,579\,038\,727\,[-5]$    & \hspace{0.5cm} & $6.438\,377\,827\,643\,[-6]$    & \hspace{0.5cm} & $7.024\,450\,825\,240\,[-6]$    \\  
          & 40  & 40  & \hspace{0.5cm} & $3.407\,420\,115\,360\,[-6]$    & \hspace{0.5cm} & $1.057\,845\,433\,238\,[-6]$    & \hspace{0.5cm} & $1.238\,406\,423\,891\,[-6]$    \\
          & 50  & 40  & \hspace{0.5cm} & $8.143\,245\,007\,846\,[-7]$    & \hspace{0.5cm} & $2.492\,961\,933\,659\,[-7]$    & \hspace{0.5cm} & $3.207\,366\,565\,785\,[-7]$    \\  
          & 60  & 50  & \hspace{0.5cm} & $2.423\,645\,144\,952\,[-7]$    & \hspace{0.5cm} & $7.284\,857\,396\,018\,[-8]$    & \hspace{0.5cm} & $1.058\,546\,736\,480\,[-7]$    \\
          & 70  & 60  & \hspace{0.5cm} & $8.294\,967\,341\,951\,[-8]$    & \hspace{0.5cm} & $2.434\,930\,605\,774\,[-8]$    & \hspace{0.5cm} & $4.127\,150\,402\,084\,[-8]$    \\  
          & 80  & 70  & \hspace{0.5cm} & $3.102\,550\,674\,083\,[-8]$    & \hspace{0.5cm} & $8.834\,009\,222\,646\,[-9]$    & \hspace{0.5cm} & $1.817\,222\,002\,459\,[-8]$    \\
          & 90  & 80  & \hspace{0.5cm} & $1.221\,173\,429\,625\,[-8]$    & \hspace{0.5cm} & $3.342\,226\,882\,557\,[-9]$    & \hspace{0.5cm} & $8.779\,997\,909\,388\,[-9]$    \\ 
\vspace{0.1cm}  
          & 100 & 100 & \hspace{0.5cm} & $4.890\,097\,303\,991\,[-9]$    & \hspace{0.5cm} & $1.269\,990\,853\,964\,[-9]$    & \hspace{0.5cm} & $2.137\,848\,312\,143\,[-9]$    \\
\hline
\hline
\end{tabular}
\end{center}
\end{table}

\twocolumngrid

\noindent defined in \Eq{r.6} are given in \tablename{~\ref{table_n2_Z1-100}}, 
and 
\beq
F(2p_{1/2},2s_{1/2}) & = & F(2s_{1/2},2p_{1/2}) \eol
& = & \dfrac{2}{9} \left\lbrace  \int_0^\infty [ P_{2p_{1/2}}(r) P_{2s_{1/2}}(r) \right. \eol
& & \left. + Q_{2p_{1/2}}(r) Q_{2s_{1/2}}(r) ] r dr \right\rbrace ^2
\eeqn{hyd:eq2}
and
\beq
F(2p_{3/2},2s_{1/2}) & = & 2 F(2s_{1/2},2p_{3/2}) \eol
& = & \dfrac{4}{9} \left\lbrace \int_0^\infty [ P_{2p_{3/2}}(r) P_{2s_{1/2}}(r) \right. \eol
& & \left. + Q_{2p_{3/2}}(r) Q_{2s_{1/2}}(r) ] r dr \right\rbrace ^2,
\eeqn{hyd:eq3}
according to Eqs. \rref{r.5} and \rref{r.6}. 
Equations \rref{hyd:eq2} and \rref{hyd:eq3} represent numerators that can be computed with the Gauss quadrature associated with the Lagrange-mesh method, 
while the denominators of Eqs.~\rref{hyd:eq1a}-\rref{hyd:eq1c} must take into account the fine-structure interval and the Lamb shift between $n=2$ states. 
\tablename{~\ref{table_other_n2_Z1}} displays the values of the numerators $F(2p_j,2s_{1/2})$ appearing in Eqs.~\rref{hyd:eq1a}-\rref{hyd:eq1c}, 
with $Z$ values between $1$ to $100$ and  given values of $N$. 
The significant digits of the results are estimated by a comparison with $N+2$ mesh points. 
For $2p_{3/2}$ one observes an accuracy of at least $10^{-13}$, with only $6$ to $10$ mesh points. 
The values for $2p_{1/2}$ are numerically exact.

\begin{table}[!ht]
\caption{Numerators of the contributions of almost degenerate states to the static dipole polarizabilities (in a.u.)
of the $n=2$ states of hydrogenic ions.
Significant digits are estimated by comparison with $N+2$ mesh points. 
Powers of $10$ are indicated within brackets.}
\label{table_other_n2_Z1}
\begin{center}
\begin{tabular}{r r c l c l}
\hline
\hline
$Z$ & $N$ & \hspace{0.25cm} & $F(2p_{1/2},2s_{1/2})$          & \hspace{0.5cm} & $F(2p_{3/2},2s_{1/2})$       \\
\hline
1   & 6   & \hspace{0.25cm} & $5.999\,733\,743\,581\,8$       & \hspace{0.5cm} & $11.999\,786\,990\,832\,3$   \\
2   & 6   & \hspace{0.25cm} & $1.499\,733\,744\,645\,1$       & \hspace{0.5cm} & $2.999\,786\,979\,582\,2$    \\
10  & 6   & \hspace{0.25cm} & $5.973\,377\,857\,887\,4\,[-2]$ & \hspace{0.5cm} & $1.197\,866\,185\,513\,[-1]$ \\
20  & 6   & \hspace{0.25cm} & $1.473\,388\,347\,402\,1\,[-2]$ & \hspace{0.5cm} & $2.978\,547\,730\,953\,[-2]$ \\
30  & 6   & \hspace{0.25cm} & $6.400\,720\,933\,264\,3\,[-3]$ & \hspace{0.5cm} & $1.311\,686\,338\,414\,[-2]$ \\
40  & 8   & \hspace{0.25cm} & $3.484\,284\,277\,272\,8\,[-3]$ & \hspace{0.5cm} & $7.280\,703\,963\,583\,[-3]$ \\
50  & 8   & \hspace{0.25cm} & $2.134\,562\,724\,010\,3\,[-3]$ & \hspace{0.5cm} & $4.576\,885\,897\,117\,[-3]$ \\
60  & 10  & \hspace{0.25cm} & $1.401\,539\,481\,366\,5\,[-3]$ & \hspace{0.5cm} & $3.105\,241\,297\,212\,[-3]$ \\
70  & 10  & \hspace{0.25cm} & $9.596\,781\,478\,383\,3\,[-4]$ & \hspace{0.5cm} & $2.214\,502\,500\,495\,[-3]$ \\
80  & 10  & \hspace{0.25cm} & $6.729\,674\,513\,475\,4\,[-4]$ & \hspace{0.5cm} & $1.632\,360\,617\,365\,[-3]$ \\
90  & 10  & \hspace{0.25cm} & $4.763\,808\,490\,110\,0\,[-4]$ & \hspace{0.5cm} & $1.228\,327\,956\,418\,[-3]$ \\
\vspace{0.1cm}
100 & 10  & \hspace{0.25cm} & $3.355\,810\,184\,607\,6\,[-4]$ & \hspace{0.5cm} & $9.330\,331\,402\,487\,[-4]$ \\
\hline
\hline
\end{tabular}
\end{center}
\end{table}

Using Eqs.~\rref{hyd:eq1a}-\rref{hyd:eq1c} requires information on energy differences. 
The corresponding transition frequencies are reviewed for hydrogen in \Ref{Kr10} 
and recent theoretical values are available on the NIST website \cite{NIST}. 
Values for hydrogenic ions can be found in \Ref{JS85}. 
As examples of the use of Eqs.~\rref{hyd:eq1a}-\rref{hyd:eq1c}, we consider two extreme cases. 

For $Z = 1$, we derive $E_{2s_{1/2}} - E_{2p_{1/2}}$ from $0.035\,285\,878(80)$ cm$^{-1}$ 
and $E_{2p_{3/2}} - E_{2s_{1/2}}$ from $0.330\,601\,966(80)$ cm$^{-1}$ \cite{NIST}. 
The $n = 2$ polarizabilities are then for hydrogen 
\beq
\bar{\alpha}_{1}^{(2s_{1/2})} &\approx& -2.935\,14 \times 10^7, 
\eoln{hyd:eq11}
\bar{\alpha}_{1}^{(2p_{1/2})} &\approx& 3.731\,79 \times 10^7, 
\eoln{hyd:eq12}
\bar{\alpha}_{1}^{(2p_{3/2})} &\approx& -3.982\,935 \times 10^6.
\eeqn{hyd:eq13}
These polarizabilities are strongly amplified by the small energy differences 
of the Lamb shift and the fine-structure interval. 
Their accuracy is limited by the accuracy on these energies. 

For $Z = 100$, we derive $E_{2s_{1/2}} - E_{2p_{1/2}}$ from $1.105 \times 10^6$ cm$^{-1}$ 
and $E_{2p_{3/2}} - E_{2s_{1/2}}$ from $5.454\,0 \times 10^7$ cm$^{-1}$ \cite{JS85}. 
The $n = 2$ polarizabilities are then 
\beq
\bar{\alpha}_{1}^{(2s_{1/2})} &\approx& -6.248 \times 10^{-5}, 
\eoln{hyd:eq14}
\bar{\alpha}_{1}^{(2p_{1/2})} &\approx& 6.734 \times 10^{-5}, 
\eoln{hyd:eq15}
\bar{\alpha}_{1}^{(2p_{3/2})} &\approx& -2.633\,3 \times 10^{-7}.
\eeqn{hyd:eq16}
They are still significantly larger than the values in \tablename{~\ref{table_n2_Z1-100}} for $2s_{1/2}$ and $2p_{1/2}$ 
but they keep the same order of magnitude for $2p_{3/2}$ where the two terms 
in \Eq{hyd:eq1c} partially cancel each other. 

As already mentioned, accurate values of polarizabilities can also easily be obtained 
for higher excited states. 
However, the increasing number of almost degenerates states requires a special treatment 
of energy differences also for higher multipolarities. 

\subsection{Yukawa potential}
\label{sec:Yukawa}

Polarizabilities can also be accurately computed for Yukawa potentials 
\beq
V(r) = -V_0\, \frac{e^{-\mu r}}{r},
\eeqn{Yuk:eq1}
with different values of $V_0$ and $\mu$. Within the Lagrange-mesh method, switching to Yukawa potentials only requires to change the potential values $V(hx_i)$ in the Hamiltonian matrix given by \Eq{Lag.7}. 
Also for this kind of potentials, it has recently been shown in \Ref{BFG14} that the Lagrange-mesh method is able to provide very accurate results with a number of mesh points for which the computation seems instantaneous. 
The approximate wave functions provide mean values of powers of the coordinate that are also extremely precise.

Potentials \rref{Yuk:eq1} have the singular behavior
\beq
V(r) \arrow{r}{0} - \frac{V_0}{r}
\eeqn{Yuk:eq2}
at the origin. 
Parameter $\gamma$ is thus given by \Eq{r.11} and parameter $\alpha$ is the same as in the Coulomb case, 
i.e., $\alpha = 2(\gamma - \vert \kappa \vert)$. 
The scaling parameter $h$ and the number $N$ of mesh points are adjusted for each potential. 

\tablename{~\ref{table_Yuk_1}} lists static dipole polarizabilities of the ground state of a hydrogen atom in a Debye plasma \cite{Da12}. 
Various values of the Debye length $\delta$ are considered. 
This situation is described by Yukawa potentials with parameter $V_0=Z=1$ and parameter $\mu=1/\delta$. 
The limit $\delta \rightarrow \infty$ corresponds to the Coulomb case. 
All computations are performed with $N=40$ mesh points and the significant digits of the results are estimated by a comparison with $N=50$. 
The scaling parameter $h$ starts from the Coulomb optimal value 0.5 and progressively increases with $\mu$. 
In the nonrelativistic case, the values are computed using Eqs.~\rref{nr.4} and \rref{nr.5}, and are compared with results reported in Refs.~\cite{QWJ09,Da12}. 
One can observe a large increase in the number of significant digits displayed with the Lagrange-mesh method, in comparison with the previous results. An accuracy of at least $10^{-12}$ is obtained with this method, while Refs.~\cite{Da12} and \cite{QWJ09} only provided up to $7$ figures. 
At the limit of the Coulomb case, $\delta \rightarrow \infty$, the exact nonrelativistic value of $4.5$ a.u.\ \cite{Bu37} is recovered with $12$ digits. 
In fact, a numerically exact value can be reached with fewer mesh points \cite{Ba12}, due to the fact that the rounding errors are increasing with $N$.

The value of the dipole polarizability increases with the screening length $\delta$, 
until reaching for $\delta=1$ a value two orders of magnitude higher than in the Coulomb limit. 
For both references \cite{Da12} and \cite{QWJ09}, the relative error is quite high. 
From $\mu=0$ to $\mu=0.5$, the Lagrange-mesh computations are closer to the results from \Ref{QWJ09} than to the ones from \Ref{Da12}. 
For larger $\mu$ values, the relative error with \Ref{Da12} stays constant, between $10^{-3}$ and $10^{-2}$, 
while the one with \Ref{QWJ09} keeps increasing, reaching $10^{-1}$ for $\mu=0.9$. 

\begin{figure}[!ht]
\begin{center}
\includegraphics[scale=0.45]{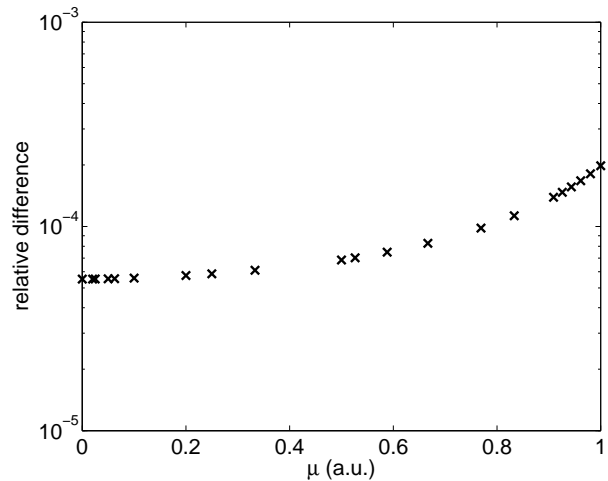}
\end{center}
\caption{Relative difference between nonrelativistic and relativistic Lagrange-mesh computations 
of the ground-state polarizability of Yukawa potentials \rref{Yuk:eq1} for $V_0=1$ as a function of $\mu$.}
\label{fig_err_Yukawa_1_R}
\end{figure}

Relativistic values are also listed in \tablename{~\ref{table_Yuk_1}}. 
They are computed using Eqs.~\rref{r.5} and \rref{r.6}. 
Relativistic values are all smaller than nonrelativistic ones, 
due to the contraction of the wave functions when relativistic effects are taken into account. 
However, this effect is small since we consider $V_0=Z=1$. 
Here also, at the Coulomb limit, the relativistic value with $N = 40$ is slightly less good than the result given in \tablename{~\ref{table_1s12_Z1_100}} with only 6 mesh points.
\figurename{~\ref{fig_err_Yukawa_1_R}} reports the relative difference between nonrelativistic and relativistic Lagrange-mesh computations, 
as a function of $\mu$. 
This difference continuously increases with $\mu$, from $5 \times 10^{-4}$ to $2 \times 10^{-3}$.

\pagebreak

\onecolumngrid

\begin{table}[!ht]
\caption{Nonrelativistic and relativistic static dipole polarizabilities (in a.u.) of the ground state of Yukawa potentials \rref{Yuk:eq1} 
for $V_0=1$ and screening lengths $\delta = 1/\mu$. 
The number of mesh points is $N=40$. 
Significant digits are estimated by comparison with $N=50$. 
Nonrelativistic values are compared with Refs.~\cite{Da12} and \cite{QWJ09}.}
\label{table_Yuk_1}
\begin{center}
\begin{tabular}{l c l c l c c c l c l}
\hline
\hline
\vspace{0.1cm}
$\delta = 1/\mu$ & \hspace{0.5cm} & \Ref{Da12}  & \hspace{0.5cm} & \Ref{QWJ09}  & \hspace{0.5cm} &        & \hspace{0.5cm} & \multicolumn{3}{c}{Lagrange-mesh ($N=40$)}                        \\
\hline
                 & \hspace{0.5cm} &             & \hspace{0.5cm} &              & \hspace{0.5cm} & $h$    & \hspace{0.5cm} & Nonrelativistic        & \hspace{0.5cm} & Relativistic            \\
$\infty$         & \hspace{0.5cm} & $4.496\,2$  & \hspace{0.5cm} & $4.500\,0$   & \hspace{0.5cm} & $0.5$  & \hspace{0.5cm} & $4.500\,000\,000\,000$ & \hspace{0.5cm} & $4.499\,751\,495\,177$  \\
50               & \hspace{0.5cm} & $4.507\,6$  & \hspace{0.5cm} & $4.508\,2$   & \hspace{0.5cm} &        & \hspace{0.5cm} & $4.508\,675\,748\,210$ & \hspace{0.5cm} & $4.508\,426\,624\,858$  \\
40               & \hspace{0.5cm} & $4.511\,9$  & \hspace{0.5cm} & $4.512\,99$  & \hspace{0.5cm} &        & \hspace{0.5cm} & $4.513\,460\,836\,744$ & \hspace{0.5cm} & $4.513\,211\,373\,377$  \\
20               & \hspace{0.5cm} & $4.550\,1$  & \hspace{0.5cm} & $4.551\,76$  & \hspace{0.5cm} &        & \hspace{0.5cm} & $4.552\,195\,439\,883$ & \hspace{0.5cm} & $4.551\,943\,236\,436$  \\
16               & \hspace{0.5cm} & $4.581\,7$  & \hspace{0.5cm} & $4.580\,03$  & \hspace{0.5cm} &        & \hspace{0.5cm} & $4.580\,489\,160\,539$ & \hspace{0.5cm} & $4.580\,234\,962\,937$  \\
10               & \hspace{0.5cm} & $4.706\,2$  & \hspace{0.5cm} & $4.699\,33$  & \hspace{0.5cm} &        & \hspace{0.5cm} & $4.699\,777\,471\,480$ & \hspace{0.5cm} & $4.699\,514\,889\,588$  \\
5                & \hspace{0.5cm} & $5.297\,6$  & \hspace{0.5cm} & $5.276\,61$  & \hspace{0.5cm} &        & \hspace{0.5cm} & $5.276\,368\,793\,394$ & \hspace{0.5cm} & $5.276\,065\,498\,754$  \\
4                & \hspace{0.5cm} & $5.770\,0$  & \hspace{0.5cm} & $5.726\,35$  & \hspace{0.5cm} &        & \hspace{0.5cm} & $5.726\,702\,088\,012$ & \hspace{0.5cm} & $5.726\,366\,480\,682$  \\
3                & \hspace{0.5cm} & $6.807\,2$  & \hspace{0.5cm} & $6.801\,56$  & \hspace{0.5cm} &        & \hspace{0.5cm} & $6.801\,959\,817\,157$ & \hspace{0.5cm} & $6.801\,544\,911\,434$  \\
2                & \hspace{0.5cm} & $11.229\,1$ & \hspace{0.5cm} & $11.147\,01$ & \hspace{0.5cm} & $0.6$  & \hspace{0.5cm} & $11.147\,655\,952\,84$ & \hspace{0.5cm} & $11.146\,892\,173\,589$ \\
1.9              & \hspace{0.5cm} & $12.390\,4$ & \hspace{0.5cm} & $12.335\,9$  & \hspace{0.5cm} &        & \hspace{0.5cm} & $12.316\,122\,276\,08$ & \hspace{0.5cm} & $12.315\,257\,969\,840$ \\
1.7              & \hspace{0.5cm} & $16.123\,5$ & \hspace{0.5cm} & $16.069\,9$  & \hspace{0.5cm} &        & \hspace{0.5cm} & $16.024\,746\,558\,53$ & \hspace{0.5cm} & $16.023\,547\,511\,631$ \\
1.5              & \hspace{0.5cm} & $24.176\,1$ & \hspace{0.5cm} & $24.098\,7$  & \hspace{0.5cm} &        & \hspace{0.5cm} & $23.952\,788\,792\,35$ & \hspace{0.5cm} & $23.950\,808\,437\,16$  \\
1.3              & \hspace{0.5cm} & $47.555\,7$ & \hspace{0.5cm} & $47.405\,2$  & \hspace{0.5cm} &        & \hspace{0.5cm} & $47.259\,844\,401\,24$ & \hspace{0.5cm} & $47.255\,206\,306\,67$  \\
1.2              & \hspace{0.5cm} & $82.219\,1$ & \hspace{0.5cm} & $82.994\,6$  & \hspace{0.5cm} &        & \hspace{0.5cm} & $81.791\,951\,912\,13$ & \hspace{0.5cm} & $81.782\,728\,920\,69$  \\
1.1              & \hspace{0.5cm} & $188.89$    & \hspace{0.5cm} & $192.910$    & \hspace{0.5cm} & $0.75$ & \hspace{0.5cm} & $187.766\,926\,629\,2$ & \hspace{0.5cm} & $187.740\,828\,417\,7$  \\
1.08             & \hspace{0.5cm} & $235.905$   & \hspace{0.5cm} & $244.785$    & \hspace{0.5cm} &        & \hspace{0.5cm} & $234.213\,472\,197\,0$ & \hspace{0.5cm} & $234.179\,068\,350\,3$  \\
1.06             & \hspace{0.5cm} & $299.589$   & \hspace{0.5cm} & $316.994$    & \hspace{0.5cm} &        & \hspace{0.5cm} & $299.883\,800\,557\,7$ & \hspace{0.5cm} & $299.836\,950\,194\,1$  \\
1.04             & \hspace{0.5cm} & $400.199$   & \hspace{0.5cm} & $409.154$    & \hspace{0.5cm} & $0.8$  & \hspace{0.5cm} & $396.221\,960\,931$    & \hspace{0.5cm} & $396.155\,620\,903$     \\
1.02             & \hspace{0.5cm} & $539.916$   & \hspace{0.5cm} & $598.717$    & \hspace{0.5cm} &        & \hspace{0.5cm} & $543.995\,546\,849$    & \hspace{0.5cm} & $543.897\,020\,777$     \\
\vspace{0.1cm}
1                & \hspace{0.5cm} & $778.723$   & \hspace{0.5cm} & $788.280$    & \hspace{0.5cm} & $0.85$ & \hspace{0.5cm} & $783.476\,574\,642$    & \hspace{0.5cm} & $783.321\,287\,554$     \\
\hline
\hline
\end{tabular}
\end{center}
\end{table}

\twocolumngrid

Relativistic polarizabilities can also be computed for other Yukawa potentials. 
These potentials were already considered in Refs.~\cite{KH02,BFG14} for the computation of the energies and mean values $\langle r^k \rangle$ of a set of bound states. 
The system of units is now $\hbar = m = c = 1$. 
\tablename{~\ref{table_Yuk_2}} reports the static dipole polarizabilities of Yukawa potentials for two cases: $\mu = 0.01$ and $V_0 = 0.1$ (corresponding to $\mu \approx 1.37$ and $V_0 \approx 13.7$ in a.u.) and $\mu = 0.04$ and $V_0 = 0.7$ (corresponding to $\mu \approx 5.48$ and $V_0 \approx 95.9$ in a.u.).

For the shallower potential, the polarizability calculations are performed with $N=40$. 
The scaling parameter $h=16$ is a good compromise for a simultaneous treatment of the three $\kappa=-1$ lowest bound states and the $\kappa=1$ and $\kappa=-2$ first bound states. 
A higher value of $22$ is needed for a better convergence of the results related to the highest two bound states, 
with the weakest binding energy and thus the largest spatial extension. 
For the deeper potential, the calculations are performed with $N = 50$. 
The scaling parameter $h = 2$ is chosen for most bound states, 
except for the ground state ($h=0.2$) which is much lower than the others (see \Ref{BFG14}) and the highest excited state ($h=5$). 
The significant digits of the results are estimated by a comparison with $N+10$ mesh points. 
For both Yukawa potentials, an accuracy of at least $10^{-12}$ is found, reaching $13$ significant figures for several bound states. 
One observes the presence of negative values of the dipole polarizability. 
According to \Eq{r.5}, the numerators in this expression are always positive. 
Thus, the minus sign may only come from the denominators of \Eq{r.5}, which contain the difference of energies between the studied state and all the  states allowed in the calculation of the dipole polarizability. 
Hence, for some bound states, the balance of all contributions can be negative.

\begin{table}[!ht]
\caption{Static dipole polarizabilities ($\hbar = m = c = 1$) of Yukawa potentials. 
Significant digits are estimated by comparison with $N+10$ mesh points. 
Powers of 10 are indicated within brackets.}
\label{table_Yuk_2}
\begin{center}
\begin{tabular}{l c r c c c c}
\hline
\hline
\vspace{0.1cm}
$n$ & \hspace{1.0cm} & $\kappa$ & \hspace{1.25cm} & $h$ & \hspace{1.25cm} & $\alpha_{1}^{(n\kappa)}$     \\
\hline
\multicolumn{7}{c}{$\mu = 0.01$, $V_0 = 0.1$ ($N = 40$)}                                                 \\
0   & \hspace{1.0cm} & $-1$     & \hspace{1.25cm} & 16  & \hspace{1.25cm} & $4.650\,527\,416\,87\,[4]$   \\
1   & \hspace{1.0cm} &          & \hspace{1.25cm} & 16  & \hspace{1.25cm} & $6.358\,097\,228\,01\,[7]$   \\ 
2   & \hspace{1.0cm} &          & \hspace{1.25cm} & 16  & \hspace{1.25cm} & $1.797\,115\,776\,56\,[9]$   \\ 
0   & \hspace{1.0cm} & $ 1$     & \hspace{1.25cm} & 16  & \hspace{1.25cm} & $-1.875\,120\,980\,415\,[7]$ \\ 
1   & \hspace{1.0cm} &          & \hspace{1.25cm} & 22  & \hspace{1.25cm} & $2.596\,925\,230\,34\,[8]$   \\ 
0   & \hspace{1.0cm} & $-2$     & \hspace{1.25cm} & 16  & \hspace{1.25cm} & $-1.733\,814\,509\,13\,[7]$  \\
\vspace{0.2cm}
1   & \hspace{1.0cm} &          & \hspace{1.25cm} & 22  & \hspace{1.25cm} & $2.868\,534\,886\,60\,[8]$   \\
\multicolumn{7}{c}{$\mu = 0.04$, $V_0 = 0.7$ ($N = 50$)}                                                 \\
0   & \hspace{1.0cm} & $-1$     & \hspace{1.25cm} & 0.2 & \hspace{1.25cm} & $9.883\,392\,685\,690$       \\
1   & \hspace{1.0cm} &          & \hspace{1.25cm} & 2   & \hspace{1.25cm} & $1.358\,955\,847\,171\,[4]$  \\ 
2   & \hspace{1.0cm} &          & \hspace{1.25cm} & 2   & \hspace{1.25cm} & $1.779\,393\,944\,896\,[5]$  \\ 
3   & \hspace{1.0cm} &          & \hspace{1.25cm} & 2   & \hspace{1.25cm} & $2.242\,455\,317\,301\,[6]$  \\ 
0   & \hspace{1.0cm} & $ 1$     & \hspace{1.25cm} & 2   & \hspace{1.25cm} & $-1.112\,555\,170\,864\,[4]$ \\ 
1   & \hspace{1.0cm} &          & \hspace{1.25cm} & 2   & \hspace{1.25cm} & $-9.103\,079\,287\,97\,[4]$  \\ 
2   & \hspace{1.0cm} &          & \hspace{1.25cm} & 2   & \hspace{1.25cm} & $-4.329\,138\,834\,76\,[5]$  \\ 
0   & \hspace{1.0cm} & $-2$     & \hspace{1.25cm} & 2   & \hspace{1.25cm} & $-1.553\,584\,893\,031\,[2]$ \\ 
1   & \hspace{1.0cm} &          & \hspace{1.25cm} & 2   & \hspace{1.25cm} & $6.686\,267\,418\,47\,[4]$   \\
\vspace{0.1cm} 
2   & \hspace{1.0cm} &          & \hspace{1.25cm} & 5   & \hspace{1.25cm} & $1.485\,180\,426\,524\,[6]$  \\
\hline
\hline
\end{tabular}
\end{center}
\end{table}


\section{Conclusion}
\label{sec:conc}
Numerically exact non-relativistic polarizabilities can be obtained with the Lagrange-mesh method \cite{Ba12}. 
This method can also provide numerically exact energies and wave functions for the Coulomb-Dirac problem. 
As shown in \Ref{BFG14}, some matrix elements are then exactly given by the associated Gauss quadrature. 
For the relativistic polarizabilities, however, the situation is more complicated. 
Partial polarizabilities with the same initial and final values of $|\kappa|$ are also exact 
but this is not the case when $|\kappa|$ varies. 
The same simple calculation then provides very accurate values with few mesh points for small charges $Z$. 
For large $Z$ values, the convergence is much slower. 
We have thus devised a new approximation, the three-mesh method, 
involving different meshes for the initial and final wave functions and for the calculation of matrix elements. 
This less elegant approach significantly improves the accuracy for high $Z$. 
The dipole polarizabilities have about thirteen significant figures agreeing 
with the highly accurate results of \Ref{TZZ12}, but are obtained with much smaller bases. 
For higher multipolarities, the high accuracies are maintained and become better than those of \Ref{TZZ12} 
for high $Z$ values. 

The simplicity of the Lagrange-mesh method allows a simple extension to the polarizabilities 
of excited states or, more precisely, to the part of these polarizabilities which does not 
involve almost degenerate states. 
We also provide the numerators of corrections allowing to include the effect of these 
almost degenerate states. 
The evaluation of polarizabilities then requires the knowledge 
of the corresponding experimental or theoretical energy differences. 
As shown by examples, for $n = 2$, these states lead to an increase of the dipole 
polarizabilities by several orders of magnitude. 
This effect decreases when $Z$ increases. 

The present approach is also valid for other potentials, with or without a singularity at the origin. 
Its efficiency and simplicity are illustrated with two Yukawa potentials. 
The first potential corresponds to a hydrogen atom in a Debye plasma. 
Relativistic and non-relativistic polarizabilities are compared with each other. 
The second Yukawa potential leads to stronger relativistic effects. 
An excellent accuracy is still obtained, even for weakly bound states. 
Properties of alkali-like atoms can easily be estimated by combining the present approach 
with the use of model and parametric potentials such as Tietz' or Green's potentials \cite{Jo07}. 

The Lagrange-mesh method is definitely accurate for estimating relativistic polarizabilities of hydrogen-like systems, 
as it was already shown for non-relativistic polarizabilities \cite{Ba12} and for relativistic energies 
and wave functions \cite{BFG14}. 
As such, Lagrange bases could find their room in the large family of finite basis sets to which B-splines and B-polynomials belong, 
to investigate two-photon processes in hydrogen-like ions \cite{AFF12,BTE14}. 

It will also be worthwhile to investigate their usefulness for the study of many-electron systems. 
The most impressive feature of the Lagrange-mesh method is the striking accuracy of the Gauss quadrature, 
that remains largely unexplained \cite{BHV02}. 
It leads in a simple way to accurate values for the nonrelativistic polarizabilities of three-body systems \cite{OB12b}. 
A more general and potentially promising route for the Lagrange-mesh method in atomic physics could be 
the use of the underlying Lagrange analytical basis that could offer interesting properties in comparison with, for instance, the B-splines. 
The latter have had a tremendous impact in atomic (relativistic) many-body calculations 
in which the pseudospectrum is constructed from B-splines confined to a large but finite cavity \cite{SJ96,SJ08}. 
The replacement of B-splines by a Sturmian basis \cite{Sz97,TKS10,BTE14} in relativistic all-order calculations of atomic properties 
\cite{SSK07} is seriously considered \cite{Sa14}. 
From this respect, we would like to point out that some Lagrange-mesh bases are exactly equivalent to Sturmian bases 
but can be simpler to use and more flexible. 


\begin{acknowledgments}
This text presents research results of the interuniversity attraction pole programme P7/12 initiated by the Belgian-state Federal Services for Scientific, Technical and Cultural Affairs. 
L.F. acknowledges the support from the FRIA.
\end{acknowledgments}


\appendix*
\renewcommand{\thesection}{\Alph{section}}
\renewcommand{\theequation}{\Alph{section}.\arabic{equation}}

\section{Equivalence between mesh expressions for polarizabilities}
\label{sec:A}

Let us start with the nonrelativistic case and show that the Lagrange-mesh approximations of Eqs.~\rref{nr.3} and \rref{nr.5} are identical. 
This property is valid because of a consistent use of Lagrange functions and Gauss quadratures. 
The coefficients $c_{nlj}$ and $c^{(1)}_{nll'j}$ of the Lagrange functions in the expansions of $\psi_{nl}$ and $\psi^{(1)}_{nll'}$ define the components of the column vectors $\ve{c}_{nl}$ and $\ve{c}_{nll'}^{(1)}$, respectively. 

Let us denote as $E_{l'}^{(k)}$ and $\ve{c}_{l'}^{(k)}$, $k = 1, \dots, N$, the eigenvalues and eigenvectors of the $N \times N$ matrix $\ve{H}_{l'}$ with elements $\la f_i | H_{l'} | f_j \ra_G$ calculated with the Gauss quadrature associated with the Lagrange mesh \cite{Ba12}, 
\beq
\ve{H}_{l'} \ve{c}_{l'}^{(k)} = E_{l'}^{(k)} \ve{c}_{l'}^{(k)}.
\eeqn{A.1}
The solution of the system corresponding to \Eq{nr.4}, 
\beq
(\ve{H}_{l'} - E_{nl} \ve{I}) \ve{c}_{nll'}^{(1)} = h^\lambda \ve{X}^\lambda \ve{c}_{nl},
\eeqn{A.2}
where $\ve{I}$ is the $N \times N$ identity matrix and $\ve{X}$ is the $N \times N$ diagonal matrix with diagonal elements $x_j$, is given by 
\beq
\ve{c}_{nll'}^{(1)} = h^\lambda (\ve{H}_{l'} - E_{nl} \ve{I})^{-1} \ve{X}^\lambda \ve{c}_{nl}.
\eeqn{A.4}
The spectral decomposition 
\beq
(\ve{H}_{l'} - E_{nl} \ve{I})^{-1} 
= \sum_{k=1}^N \ve{c}_{l'}^{(k)} (E_{l'}^{(k)} - E_{nl})^{-1} \ve{c}_{l'}^{(k)T}
\eeqn{A.5}
exactly transforms the Lagrange-mesh approximation of \Eq{nr.5},
\beq
\alpha^{(nll')}_{\lambda} = 2 (2l'+1) 
\left ( \begin{array}{ccc} l' & \lambda & l \\  0 & 0 & 0 \end{array} \right )^2  
h^\lambda \sum_{j=1}^N c_{nll'j}^{(1)} x_j^\lambda c_{nlj}, \eol
\eeqn{A.6}
into the Lagrange-mesh approximation of \Eq{nr.3}, 
\beq
\alpha^{(nll')}_{\lambda} & = & 2 (2l'+1) 
\left ( \begin{array}{ccc} l' & \lambda & l \\  0 & 0 & 0 \end{array} \right )^2 \eol
& & \times h^{2\lambda} \sum_{k=1}^N \frac{(\sum_{j=1}^N c_{l'j}^{(k)} x_j^\lambda c_{nlj})^2}{E_{l'}^{(k)} - E_{nl}}.
\eeqn{A.7}
In the Coulomb case, this equivalence remains true in the presence of a degenerate level 
by eliminating the $k = n - l'$ term in Eqs.~\rref{A.5} and \rref{A.7}. 

The same proof holds in the relativistic case for the equivalence of the Lagrange-mesh approximations of Eqs.~\rref{r.5} and \rref{r.8}. 
Matrix $\ve{H}_{l'}$ must be replaced by the $2N \times 2N$ matrix $\ve{H}_{\kappa'}$ and vectors $\ve{c}_{nl}$, $\ve{c}_{nll'}^{(1)}$, and $\ve{c}_{l'}^{(k)}$ must be replaced by vectors containing the $2N$ coefficients of the Lagrange functions in the corresponding expansions of the large and small components. 


\section{Computation of Lagrange functions}
\label{sec:B}

Numerical values of a Lagrange function $\hat{f}_j^{(\alpha)}(x)$ can be computed with \Eq{Lag.3} but this expression becomes inaccurate when $x$ is close to $x_j$ because the numerator and denominator simultaneously vanish. 
In a small interval $(x_j - \epsilon, x_j + \epsilon)$, expression \rref{Lag.3} can be approximated by its second-order Taylor expansion around $x_j$, 
\beq
& & \hat{f}_j^{(\alpha)}(x) \approx \lambda_j^{-1/2} 
\left\{ 1 + \frac{x-x_j}{2x_j} \right. \eol
& & \left. - \frac{[(4N+2\alpha+2-x_j)x_j+4-\alpha^2](x-x_j)^2}{24x_j^2} \right\}. \eol
\eeqn{B.1}
With $\epsilon \approx 10^{-4}$, one obtains at least an eleven-digit accuracy everywhere. 

More precise values can be obtained with the alternative expression of the Lagrange functions \cite{BH86} 
\beq
\hat{f}_j^{(\alpha)}(x) = \sqrt{\lambda_j}\, \frac{x}{x_j} 
\sum_{k=1}^{N}\varphi_k^{(\alpha)} (x_j) \varphi_k^{(\alpha)} (x) 
\eeqn{B.2}
with 
\beq
\varphi_k^{(\alpha)} (x) = \left(\frac{k!}{\Gamma(\alpha + k + 1)}\right)^{1/2} 
x^{\alpha/2} e^{-x/2} L_k^\alpha(x). \eol
\eeqn{B.3}
This calculation provides a uniform accuracy over the whole interval but requires many more evaluations of Laguerre polynomials.


\end{document}